\documentclass[useAMS,usenatbib]{mn2e}

\usepackage{graphicx}
\usepackage{natbib}

\newcommand\ion[2]{\mbox{#1\,{\sc #2}}}
\newcommand\aap{A\&A}
\newcommand\pasp{PASP}
\newcommand\aj{AJ}
\newcommand\mnras{MNRAS}
\newcommand\apj{ApJ}
\newcommand\apjs{ApJS}
\newcommand\aaps{A\&AS}
\newcommand\ibvs{IBVS}

\title[VV Orionis]{Observational studies of early-type binary stars:  VV~Orionis}
\author[Terrell, \it{et al.}]
{Dirk Terrell$^{1}$\thanks{E-mail: terrell@boulder.swri.edu (DT); ulisse.munari@oapd.inaf.it (UM); siviero@pd.astro.it (AS)},
Ulisse Munari$^{2}$ and Alessandro Siviero$^{3}$\\
$^{1}$Department of Space Studies, Southwest Research Institute, 1050 Walnut St., Suite 400, Boulder, CO 80302, USA\\
$^{2}$Osservatorio Astronomico di Padova, Sede di Asiago, I-36032 Asiago (VI), Italy\\
$^{3}$Osservatorio Astronomico di Padova, Sede di Asiago, I-36032 Asiago (VI), Italy\\}
\begin{document}

\date{Accepted 2006 August 3. Received 2006 August 3; in original form 2006 August 3}

\pagerange{\pageref{firstpage}--\pageref{lastpage}} \pubyear{2006}

\maketitle

\label{firstpage}

\begin{abstract}

New and previously published observations of the bright eclipsing binary VV~Orionis are
analyzed.  We present new radial velocities and interstellar reddening measurements from
high-resolution spectra of this detached, short-period (P=1.48 d) binary. We discuss the
validity of prior claims for the existence of a third body and show that our new velocities
and light curve solution cast doubt on them. The components of VV~Ori are shown to be a B1 V
primary with a mass $M{_1}=10.9 \pm 0.1 M_{\sun}$ and a radius $R_{1}=4.98 \pm 0.02 R_{\sun}$ 
and a B4.5 V secondary with a mass $M{_2}=4.09 \pm 0.05 M_{\sun}$ and a radius 
$R_{2}=2.41 \pm 0.01 R_{\sun}$.

\end{abstract}


\section{Introduction}

VV Orionis (HD 36695) is a bright ($V$=5.4), double-lined eclipsing binary consisting of
main sequence B-type stars in a detached configuration. The inclination is high enough to 
produce complete eclipses, enabling the very accurate determination of fundamental 
parameters of the system. Since the discovery of its photometric variability over a century ago, 
the system has been the subject of several spectroscopic and photometric studies as
discussed by \cite{sv95}. Unfortunately, the derived absolute parameters in these previous studies
vary considerably and this led us to obtain additional spectroscopic and photometric data on
the system as part of our program on early-type binaries (viz. \cite{ter03}, \cite{ter05}).

Dating back to the first comprehensive spectroscopic study by \cite{daniel15}, spectroscopic
investigators, including \cite{sl49}, \cite{bg69}, \cite{duer75} and \cite{popper93}, have concluded that
a third star orbits the eclipsing pair. The evidence for claims of this third body rests solely
on higher than expected residuals in fits to the radial velocities of the primary as the third body's
lines are not seen in the system's spectrum. \cite{daniel15} offered the third body hypothesis "with 
great reserve" and \cite{sl49} pointed out that the observed variation in the systemic
velocity did not necessarily imply that the system was a triple. \cite{duer75} analyzed his 
radial velocities along with those of \cite{daniel15}, \cite{sl49} and \cite{bg69} to
arrive at orbital parameters for a third body orbiting with a period of about 119 days.
A goal of the current study was to investigate this third body hypothesis.

\section{Observations}
\subsection{Photometry}

VV Ori was observed with a 0.25m Schmidt-Cassegrain telescope, Santa Barbara Instrument Group ST-7XE CCD 
camera and standard $BVI_C$ filters. The main goal of the new photometric observations was to cover at
least one primary eclipse for the purpose of extending the baseline of eclipse timings so that any small
changes in the period could be detected. Given the extensive photometric datasets previously published,
we did not attempt to observe complete light curves for the system.

The CCD images were reduced using IRAF. Bias and dark frame subtraction and flatfielding were done in the
usual way. PyRAF scripts automated the frame calibrations and aperture phometry which was performed using 
the IRAF DAOFIND and PHOT routines. The comparison star for the differential photometry was
HD 36779.

\subsection{Spectroscopy}

High resolution, high signal-to-noise spectra of VV~Ori were obtained with the Echelle+CCD 
spectrograph of the 1.82m telescope operated in Asiago by the INAF Astronomical 
Observatory of Padova. The CCD is an E2V thinned, back illuminated 
1K$\times$1K chip of 13.5 $\mu$m pixel size. The slit was always aligned with 
the parallactic angle and its projected width was 1.5 arcsec. The spectra cover 
the range from atmospheric cut-off around 3250\AA\ to 7100\AA\ at a resolving 
power 30,000. The accuracy of the zero point of the wavelength scale is derived on each 
observed spectrum by cross-correlation against a synthetic template of the rich 
telluric absorption spectrum at wavelengths 6860-7050\AA, 6270-6310\AA, 5880-5950\AA\ 
and found to be affected by no systematic shift larger than 0.2 km sec$^{-1}$, 
consistent with the high efficiency of the instrument in deriving accurate radial 
velocities evaluated by \cite{ml92}. The exposure times for the spectra were 5 minutes (0.002 in 
orbital phase), eliminating any concerns about phase smearing effects on the radial velocities. 
In all, 31 spectra were obtained over the period from November, 2004 to February, 2006.

\begin{figure}
 \includegraphics[width=84mm]{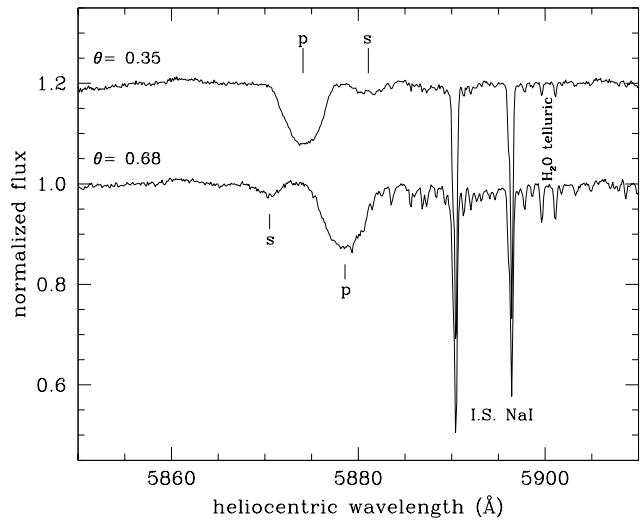}
 \caption{The appearance of the \ion{He}{i} 5876~\AA\ line in the spectrum of VV~Ori. The line components
  originating from the primary and secondary star are labeled "p" and "s", respectively, and $\theta$
  is the orbital phase. The interstellar sodium doublet and the telluric H$_2$O 
  absorption lines longward of 5885~\AA\ are also marked.} 
 \label{heplot}
\end{figure}

Figure \ref{heplot} shows the appearance of the \ion{He}{i} 5876\AA\ line near quadratures. The weakness of the
secondary's lines illustrates the need for high-resolution, high signal-to-noise observations when
trying to measure the radial velocity of the secondary. Because of the weakness of the secondary's 
lines, its radial velocities could only be accurately measured near quadratures when the blending
with the primary's lines was minimized.

\subsubsection{Reddening}

The \ion{Na}{i} lines have two interstellar components with radial velocities
of +9.3$\pm$0.2 and +26.1$\pm$0.1 km sec$^{-1}$. The FWHM of both components
is 14.4 km/sec, which is dominated by the instrumental PSF. The equivalent
widths of the two components are 0.066$\pm$0.003 and 0.160$\pm$0.006~\AA\ 
for the \ion{Na}{i} D2 line at 5890\AA, and 0.040$\pm$0.002 and 0.130$\pm$0.005~\AA\ 
for the \ion{Na}{i} D1 line at 5896\AA. Using the \cite{mun97} calibration on NaI line 
at 5890 \AA\, this corresponds to a reddening of E(B-V)=0.078$\pm$0.004.

\begin{figure}
 \includegraphics[width=84mm]{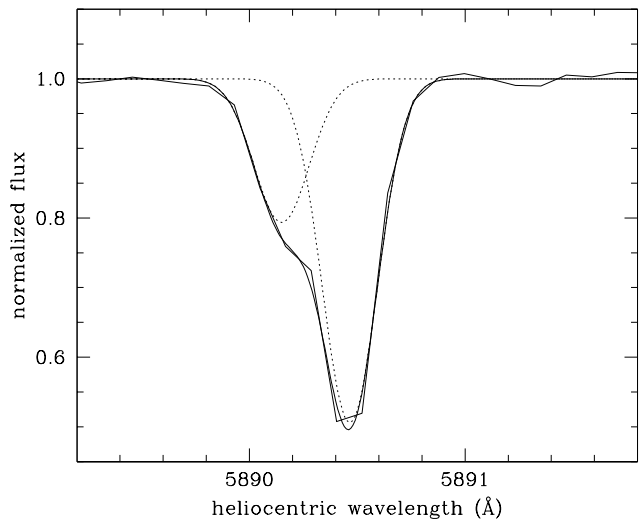}
 \caption{The interstellar line of \ion{Na}{i} D2 with the two-component 
Gaussian fit used in the reddening determination. The thick line is the observed profile
and the thin line shows the two-component fit. The dashed lines show the individual
components of the fit.} 
 \label{naiplot}
\end{figure}

The Tycho $B_T - V_T = -0.20 \pm 0.01$ corresponds, according to \cite{b00}, to a Johnson
$B-V = 0.17 \pm 0.01$. We denote an intrinsic (\it i.e.\rm, unreddened) color by $(B-V)^\prime$ and an 
observed color by $(B-V)$ with subscripts $1$, $2$ and $C$ denoting the primary 
component, the secondary component and the composite value for the binary respectively.
With an assumed intrinsic color
for the primary, $(B-V){}_1^\prime$, and the magnitude differences between the two components
in each filter, $\Delta m_B$ and $\Delta m_V$, from our light curve solution,
we can compute the intrinsic composite color of the binary, $(B-V){}_C^\prime$, as
\begin{displaymath}
    (B-V){}_C^\prime=(B-V){}_1^\prime - 2.5\; \log{
                               \left( \frac{1+10^{(-0.4\times\Delta m_B)}}
                                    {1+10^{(-0.4\times\Delta m_V)}} \right) 
                              }
\end{displaymath} 
Assuming a value of $(B-V)_1^\prime = -0.27$ for the B1V primary \citep{straizys92} and the $\Delta m_B$ and
$\Delta m_V$ values of 2.58 and 2.49 respectively from the light curve solution, we find
$(B-V){}_C^\prime = -0.26$. This result, combined with the observed $(B-V)_C$, yields a reddening 
of $E(B-V)=0.09$, in good agreement with the results from the analysis of the interstellar 
\ion{Na}{i} lines. These values also agree well with the value of 0.07 measured by both \cite{duer75}
and \cite{cl82}.

\noindent

\subsubsection{Radial velocities}

Radial velocity measurements for the primary have been derived from measurements 
of seven hydrogen Balmer lines ($\alpha$, $\beta$, $\gamma$, $\delta$, $\epsilon$, 
H8, H9) and eight \ion{He}{I} lines (7065, 5876, 5016, 4472, 4388, 4026, 4009, 3820 \AA).

Given the weakness of its spectral features (cf. Figure \ref{heplot}), the measurement of the 
radial velocity of the secondary star has been possible only in a few spectra, those 
with the highest S/N near quadrature, and for only a subset of the \ion{He}{I} lines. Table~2 
gives the radial velocities of both components.

\begin{table}
 \label{rvtable}
  \caption{Heliocentric radial velocities of VV~Ori.}
  \begin{tabular}{ccc}
\hline
HJD & Primary R.V. & Secondary R.V.\\
(mid-exposure) & (km sec$^{-1}$) & (km sec$^{-1}$)\\
\hline
2453314.5470 & -88.29 & \\
2453373.4771 &  97.01 & \\
2453373.4804 &  93.64 & \\
2453373.4828 &  95.12 & \\
2453392.4492 & 140.27 & -241.88   \\
2453392.4538 & 142.14 & -241.59  \\
2453392.4581 & 142.12 & -246.59  \\
2453393.4475 & -77.80 &  288.75  \\
2453393.4521 & -76.09 &  275.65  \\
2453393.4565 & -72.99 &  269.44  \\
2453394.3750 &  52.35 & \\
2453394.3794 &  47.78 & \\
2453394.3841 &  47.40 & \\
2453411.3610 & -37.56 & \\
2453411.3689 & -34.81 & \\
2453411.3767 & -32.80 & \\
2453411.4244 & -11.41 & \\
2453411.4323 & -7.48  & \\
2453411.4400 & -7.59  & \\
2453412.3884 & -58.46 &  \\
2453412.3963 & -57.64 &  \\
2453412.4044 & -63.07 &  \\
2453775.2399 & -59.49 &  \\
2453777.3065 & 144.07 & -274.19 \\ 
2453777.3125 & 144.42 & -282.39  \\
2453777.3389 & 142.22 & -279.10  \\
2453780.2507 & 147.63 &   \\
2453780.2551 & 147.51 &   \\
2453780.3954 & 118.14 & -234.42\\
2453780.4003 & 114.54 & -230.26\\
2453780.4050 & 112.87 & -226.48\\
\hline
\end{tabular}
\end{table}

\section{Data Analysis}
\label{analysis}

We performed a simultaneous analysis of the \cite{cl82} $UBV$ and $vby$ photometry,
our $UBV$ photometry and our radial velocities with the 2003 version of the Wilson-Devinney 
(\citealt{wd71}; \citealt{rew79}; \citealt{rew90};hereafter, WD) program. We originally
included the $u$ observations of \cite{cl82} but found, as they did, that we could not get
satisfactory fits to those data. Inspection of the light curves shows
that the system is detached so we used WD mode 2 in the solution. The parameters adjusted in the simultaneous solution were 
the semi-major axis of the relative orbit ($a$), the binary centre of mass radial velocity ($V_{\gamma}$), 
orbital inclination ($i$), secondary mean effective temperature ($T_2$), modified surface potentials of the 
components ($\Omega_1$ and $\Omega_2$), mass ratio ($q$), and the bandpass-specific luminosity of the primary 
($L_1$). We used time as the independent variable and adjusted the orbital period ($P$), and the reference epoch ($HJD_0$).
No evidence for an orbital eccentricity was found in light curve fits so we fixed it at zero for the final solution. 
Certain parameters, such as the bolometric albedos and gravity brightening exponents, were held fixed at their expected 
theoretical values. The logarithmic limb darkening law was used with coefficients from \cite{wvh93}. 
The rotational velocities of the primary (167 $\pm$ 4~km~sec$^{-1}$) and the secondary (83 $\pm$ 8~km~sec$^{-1}$) are
consistent with the assumption of synchronous rotation and that was assumed in the solution. 
The mean effective temperature of the primary was set to 26199 K based on the unreddened $B-V$ 
and the calibration of \cite{pjf96}. Kurucz atmosphere models \citep{k93} were used for both stars. Data set 
weights were determined by the scatter of the observations. 

Given the frequent claims of the existence of a third body in the system, we also adjusted third light
but we could not find statistically significant values for any of the light curves. \cite{cl82} also found
no indication of third light in their data. \cite{sv95} analyzed the $UBV$ data of \cite{duer75} and the
$H_\alpha$ data of \cite{cd87} and also found no evidence of third light. \cite{chambliss84}, however, did find
a third light contribution but did not give error estimates on which to judge the significance of the result.

\begin{figure}
 \includegraphics[width=84mm]{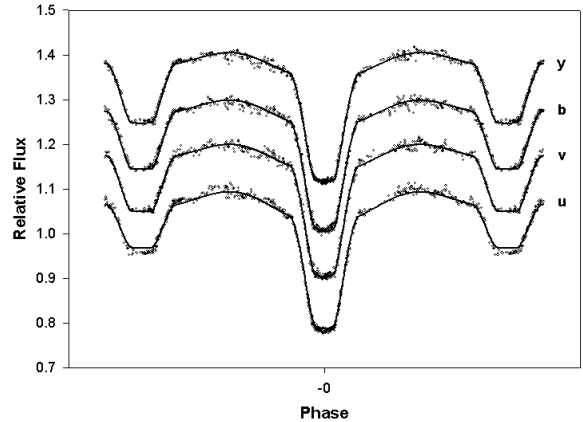}
 \caption{The $uvby$ photometry from Chambliss and Leung (1982) and the computed light curves. The $u$ data
   were not used in the final solution and are shown only to illustrate the nature of the fitting difficulties.}
 \label{vbyfits}
\end{figure}

\begin{figure}
\includegraphics[width=84mm]{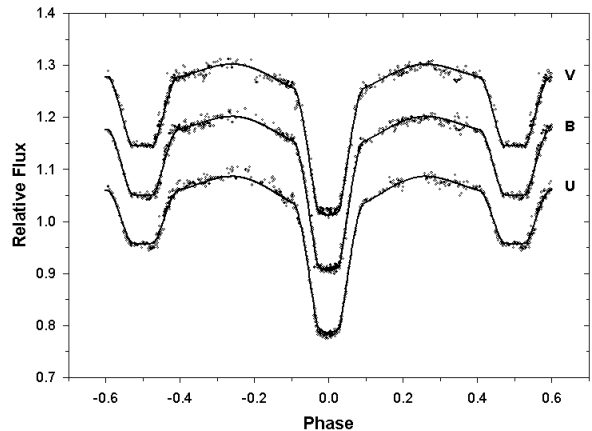}
\caption{The $UBV$ photometry from Chambliss and Leung (1982) and the computed light curves.}
\label{ubvfits}
\end{figure}

\begin{figure}
\includegraphics[width=84mm]{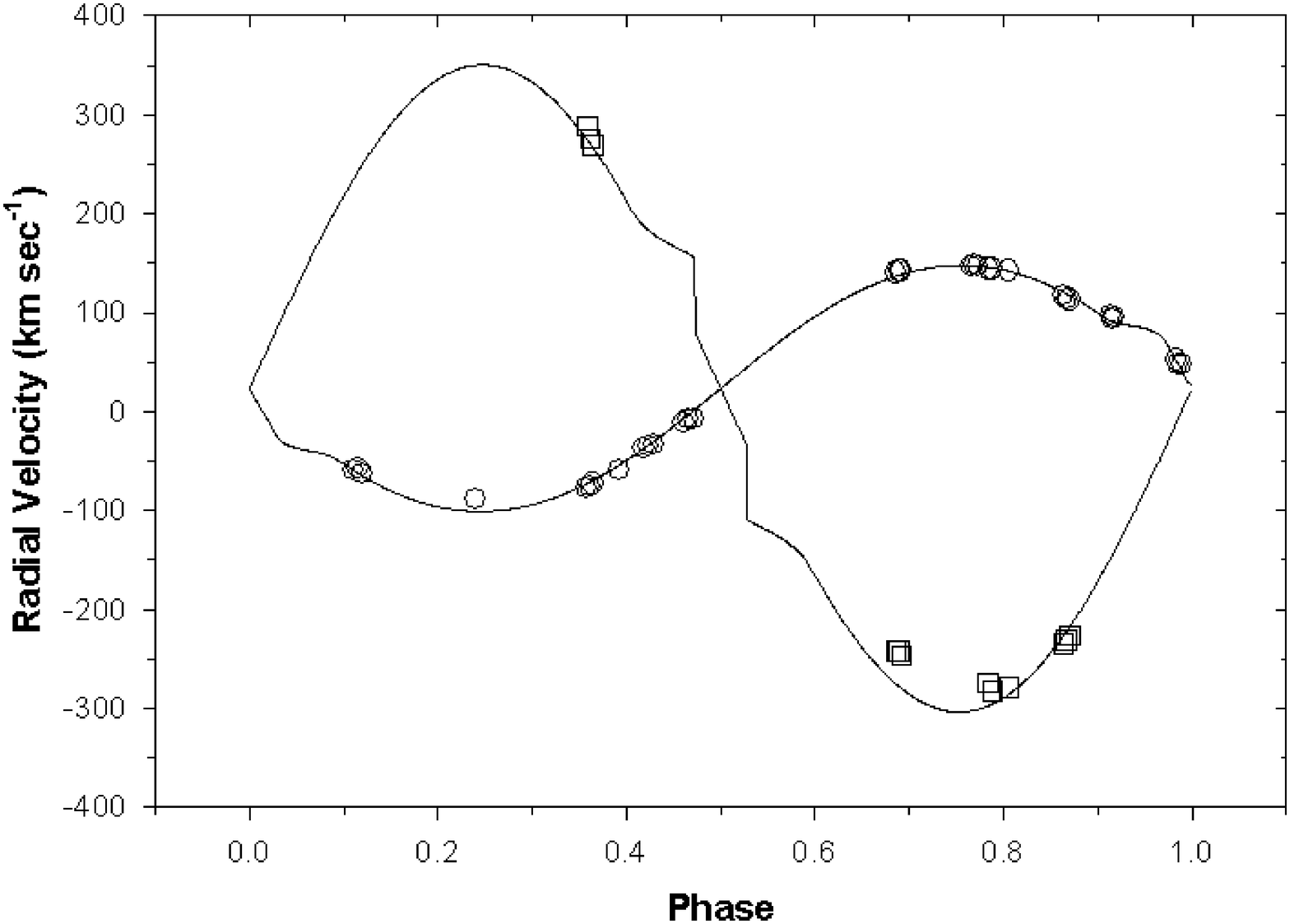}
\caption{The radial velocities from the current study and the computed curves.}
\label{rvfit}
\end{figure}

Table \ref{parameters} shows the results of 
the simultaneous solution and Figures \ref{vbyfits} and \ref{ubvfits} show the 
fits to the \cite{cl82} data. Figure \ref{rvfit} shows the fits to the radial velocities. The primary
appears to have absolute dimensions consistent with a slightly evolved main sequence star of its spectral type. 
Using a stellar evolution code (\cite{han94}, \cite{ppe73}, \cite{ppe72}, \cite{ppe71}) kindly 
supplied to us by P. Eggleton, we find that the VV~Ori primary reaches its current radius at
an age of 8.3 Myr, with the error in the radius leading to an uncertainty in the age of 
0.2 Myr. At that age, the models predict a secondary radius of 2.44 $R_{\sun}$ in reasonably good agreement with 
our observed value of 2.41 $R_{\sun}$ given the uncertainties in the masses. \cite{cl82} found that 
the secondary was much smaller than expected for the mass they derived. Our derived mass is 
considerably smaller and our radius slightly larger than their values, easing this mass-radius 
discrepancy. Table \ref{absparcomp} shows our results for the component masses and radii compared with previously 
published values. 

\cite{popper93} estimates the spectral type of the secondary as B5 V. Our derived effective temperature
results in a B4.5 classification according to the tables of \cite{straizys92} and \cite{bert04}. The
radius of 2.41 R$_{\sun}$ is consistent with an unevolved B4 V star \citep{straizys92} and the surface 
gravity is also consistent with that spectral type. The secondary's bolometric magnitude is -1.57
and that value is in good agreement with the \cite{straizys92} tables for a B4.5 V star. Experiments with composite 
reference spectra and the strength of the \ion{He}{I} 5876 line both point to a B4.5 V spectral 
classification. Based on these factors, we estimate that the secondary's spectral type is B4.5 V.

\begin{table}
  \caption{Parameters of VV Ori}
  \label{parameters}
  \begin{tabular}{cc}
  \hline
   Parameter & Value\\
  \hline
  $a$ & $13.49 \pm 0.05 R_{\sun}$  \\
  $V_{\gamma}$ & $23.5 \pm 0.3$ km sec$^{-1}$\\
  $i$ & $85^{\circ}.9 \pm 0^{\circ}.2$  \\
  $T_{1}$ & $26199$ K \\
  $T_{2}$ & $16073 \pm 42$ K\\
  $\Omega_{1}$ & $3.152 \pm 0.004$ \\
  $\Omega_{2}$ & $3.433 \pm 0.008$ \\
  $q$ &  $0.376 \pm 0.001$ \\
  $HJD_{0}$ &  $2452500.2065 \pm 0.0003$ \\
  $P$ & $1.48537423 \pm 0.00000005$ days \\
  $L_{1}/(L_{1}+L_{2})_U$ & $0.9367 \pm 0.0005$  \\
  $L_{1}/(L_{1}+L_{2})_B$ & $0.9150 \pm 0.0005$  \\
  $L_{1}/(L_{1}+L_{2})_V$ & $0.9083 \pm 0.0005$  \\
  $L_{1}/(L_{1}+L_{2})_{I_C}$ & $0.8999 \pm 0.0014$  \\
  $L_{1}/(L_{1}+L_{2})_v$ & $0.9165 \pm 0.0005$  \\
  $L_{1}/(L_{1}+L_{2})_b$ & $0.9111 \pm 0.0005$  \\
  $L_{1}/(L_{1}+L_{2})_y$ & $0.9081 \pm 0.0004$  \\
  $R_{1}$ & $4.98 \pm 0.02$ R$_{\sun}$ \\
  $R_{2}$ & $2.41 \pm 0.01$ R$_{\sun}$ \\
  $M_{1}$ & $10.9 \pm 0.1$ M$_{\sun}$ \\
  $M_{2}$ & $4.09 \pm 0.05$ M$_{\sun}$ \\
  $L_{1}$ & $10600 \pm 1600$ L$_{\sun}$\\
  $L_{2}$ & $350 \pm 54$ L$_{\sun}$\\
  $log~g_{1}$ & $4.08 \pm 0.06$  \\
  $log~g_{2}$ & $4.29 \pm 0.06$ \\
  \hline
  \end{tabular}

\medskip
Quoted errors are formal 1$\sigma$ errors from the solution.
Luminosity errors are estimates based on an estimated uncertainty of 1000 K
in the effective temperature of the primary. The $log~g$ values are in CGS units.
\end{table}

\begin{table*}
  \caption{Comparison of measurements of the masses and radii of the components of VV Ori.}
  \label{absparcomp}
  \begin{tabular}{cccccc}
  \hline
   Parameter & Current Study & Popper (1993) & Chambliss \& Leung (1982) & Duerbeck (1975) & Eaton (1975)\\
  \hline
  $R_1$ ($R_{\sun}$) & $4.98 \pm 0.03$ & $5.0 \pm 0.2$  & 4.94 & 4.47 & 5.0\\
  $R_2$ ($R_{\sun}$) & $2.41 \pm 0.02$ & $2.45 \pm 0.1$ & 2.36 & 2.21 & 2.3\\
  $M_1$ ($M_{\sun}$) & $10.9 \pm 0.2$  & $10.8 \pm 0.6$ & 10.2 & 7.60 & 9.3\\
  $M_2$ ($M_{\sun}$) & $4.09 \pm 0.08$ & $4.5 \pm 0.8$  & 4.5  & 3.42 & 4.3\\
  \hline
  \end{tabular}
\end{table*}

Assuming a ratio of total to selective absorption of $R=3.1$ and using the bolometric corrections
from \cite{pjf96}, we find that the distance to VV~Ori is 388 $\pm$ 30 parsecs with the error
dominated by the uncertainty in the effective temperature of the primary. Unfortunately,
the Hipparcos parallax is not very precise, yielding a $1\sigma$ distance range of 390pc to 1060 pc.
Our distance places the system in the Orion OB1 association consistent with the distance to the 1a 
subgroup \citep{blaauw64} measured by \cite{dezeeuw99}. \cite{sv95} estimate that the age of VV~Ori 
is 10 $\pm$ 1 Myr. This value, along with the age found from our evolutionary calculations, supports 
the membership of VV~Ori in the 1a subgroup of Orion OB1 whose age was measured to be 11.4 $\pm$ 1.9 Myr.
The barycentric radial velocity we measure, $23.5 \pm 0.5$ km sec$^{-1}$, agrees well with the average 
radial velocity of the Ori OB1 members found by \cite{ml91}.At the galactic latitude of $-18^\circ$ 
and our distance, VV~Ori is 110 pc from the galactic plane, within the scale height of the galactic 
thin disk.

\section{The Putative Tertiary Component}

\cite{daniel15} was the first to suggest the existence of a tertiary component in VV~Ori.
Larger than expected residuals from the fit to his radial velocity observations of the primary
led him to examine the possibility of a third body and he gave orbital parameters for it
"with great reserve ... and not with any idea that they are even approximately true."
Later, \cite{sl49} also noted systematic trends in the residuals to the fits of their
velocities but did not feel justified in solving for the orbit of a third body, stating that
"... we cannot be certain that the slow variation necessarily means that the system
is triple." \cite{duer75} analyzed the velocity residuals for the \cite{daniel15}, 
\cite{sl49} and \cite{bg70} data sets, as well as his own, concluding that a third
body orbit with an period of about 119 days fit the data best. Assuming an inclination
of $90^{\circ}$ for the orbit of the third body, he estimated that its spectral type
was mid-to-late A and its luminosity was 13-14 $L_{\sun}$.

\begin{figure}
\includegraphics[width=84mm]{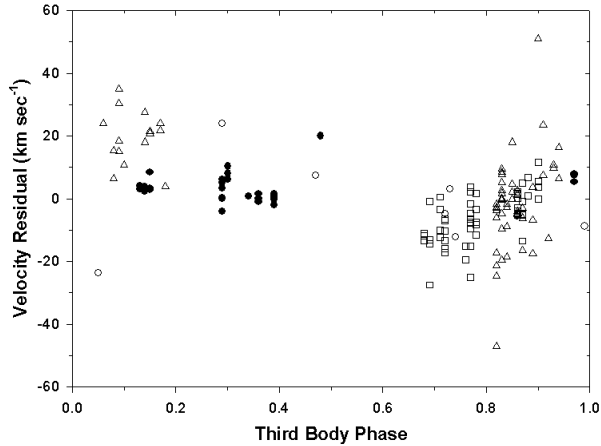}
\caption{The radial velocity residuals for the simultaneous fit to several data sets, phased with the
third body ephemeris given by Duerbeck (1975).  Triangles
are from the observations of Struve and Luyten(1949), squares from Duerbeck (1975) and open circles from 
Popper (1993). Data from the current study are shown with filled circles.}
\label{residualfit}
\end{figure}

We repeated our light and velocity curve analysis after adding the velocities from 
\cite{popper93}, \cite{duer75} and \cite{sl49} to study the behaviour of the velocity residuals
for each radial velocity data set. All radial velocities were given equal weight in the solution.
Figure \ref{residualfit} shows the velocity residuals for the four sets of data phased with 
the third body ephemeris given by \cite{duer75}. While the older data appear to show systematic
trends, our more precise data do not, but we, unfortunately, do not have full coverage of the
purported third body orbit. 

One intriguing result of the analysis is a statistically 
significant value for the time derivative of the orbital period of the binary. The solution
converged on a value of $\dot{P} = -3.6 \times 10^{-10} \pm 1.3 \times 10^{-10}$, perhaps
indicating accretion of the primary's wind by the secondary. We do not attach much significance
to this result, its reality having to be assessed against additional precise radial
velocities and accurate timing of eclipses to be secured in the years to come.
In the case of conservative mass transfer between the two components, the mass 
exchange would amount to $2 \times 10^{-7} M_{\sun}~yr^{-1}$. This material would 
be easily ionized by the radiation field of the B1 V primary, and 
should give rise to emission lines easily detectable in high 
resolution spectra. Ours do not show any hint of emission lines originating in VV Ori

Another way to test for the existence of the third body, as pointed out by \cite{duer75},
is through its effect on the O-C diagram for times of minimum of the binary. Unfortunately, VV~Ori has
not been a frequent target of observers of times of minimum. A concerted effort to 
observe as many times of minimum as possible over a couple of observing seasons could
prove valuable in answering this critical question. \cite{duer75} predicts that the 
double amplitude of the variation would be 0.0014 days, a measurement within reach
of a small telescope equipped with a CCD camera.

Given the large scatter in the older radial velocity data sets, the lack of any evidence for third
light in the light curves and the lack of systematic trends in the residuals of our radial
velocities, we see no strong evidence to support claims of the existence of a third body in VV~Ori. 
More extensive radial velocity and time of minimum studies should be able to resolve the question
with greater certainty.

\section{Conclusions}

VV Ori consists of a B1 V primary with $M{_1}=10.9 \pm 0.1 M_{\sun}$ and $R_{1}=4.98 \pm 0.02 R_{\sun}$ 
and a B4.5 V secondary with $M{_2}=4.09 \pm 0.05 M_{\sun}$ and $R_{2}=2.41 \pm 0.01 R_{\sun}$. The
derived distance of 388 $\pm$ 30 pc and barycentric velocity of $23.5 \pm 0.3$ km sec$^{-1}$ are consistent 
with the system's membership in the Orion OB1 association. We find no evidence of third light in the 
photometric data and the new radial velocities, although somewhat limited in temporal coverage, show no 
evidence of the third body discussed by previous authors. 

\section*{Acknowledgments}

This research has made use of the SIMBAD database, operated at CDS, Strasbourg, France.
DT acknowledges the support of the American Astronomical Society through its Small 
Research Grants Program. We thank the referee, Helmut A. Abt, for a constructive review 
of the manuscript that helped clarify several points.

\bsp

\label{lastpage}

\end{document}